\documentclass[12pt]{article}
\usepackage{graphicx,amsmath,amsfonts}

\def\NC   {\rm Nuovo Cimento }

\def\NIMA{{\rm Nucl. Instr. and Meth.} {\bf A}}

\def\NPB {{\rm Nucl. Phys.} {\bf B}}
\def\PLB {{\rm Phys. Lett.} {\bf B}}
\def\PRL  {\rm Phys. Rev. Lett. }
\def\PR  {{\rm Phys. Rev. }}
\def\PRC {{\rm Phys. Rev.} {\bf C}}
\def\PRD {{\rm Phys. Rev.} {\bf D}}

\def\RMP  {\rm Rev. Mod. Phys. }

\newcommand{\etal}{{\em et al.}}

\def\etap{\eta^{\prime}}

\def\ba{\begin{eqnarray}}\def\ea{\end{eqnarray}}
\def\bc{\begin{center}}\def\ec{\end{center}}
\def\nn{\nonumber\\}

\title{Photoproduction of Pseudoscalar Mesons off Nuclei at Forward Angles}

\author{ 
 S.~Gevorkyan\footnote{Corresponding author:gevs@jlab.org.
 On leave of absence from Yerevan Physics Institute} \\
 Joint Institute for Nuclear Research \\ Dubna, Moscow region, 141980, Russia 
\and
 A.~Gasparian \\
 North Carolina A\&T State University \\ Greensboro, NC 27411, USA 
\and
 L.~Gan \\
 University of North Carolina Wilmington \\ Wilmington, NC 28403, USA
\and
 I.~Larin \\
 Institute for Theoretical and Experimental Physics \\ Moscow, Russia
\and
 M.~Khandaker \\
 Norfolk State University \\ Norfolk, VA 23504, USA
}

\begin{document}

\maketitle

\begin{abstract}

With the advent of new photon tagging facilities and novel
experimental technologies it has become possible to perform
photoproduction cross section measurements of pseudoscalar mesons on
nuclei with a percent level accuracy.  The extraction of the radiative
decay widths from these measurements at forward angles is done by the
Primakoff method, which requires theoretical treatment of all
processes participating in these reactions at the same percent level.
In this work we review the theoretical approach to meson
photoproduction amplitudes in the electromagnetic and strong fields of
nuclei at forward direction.  The most updated description of these
processes are presented based on the Glauber theory of multiple
scattering.  In particular, the effects of final state interactions,
corrections for light nuclei, and photon shadowing in nuclei are
discussed.
\end{abstract}

\section{Introduction}

The properties of QCD at low energies are manifested in their most
unambiguous form in the sector of light pseudoscalar mesons $\pi^0$,
$\eta$ and $\etap$.  The two-photon decays of these mesons are
primarily caused by the chiral anomaly~\cite{BellJackiw69,Adler69},
the explicit breaking of a classical symmetry by the quantum
fluctuations of the quark fields when they couple to the
electromagnetic field.  This anomalous symmetry breaking is of a pure
quantum mechanical origin and can be calculated exactly to all orders
in the chiral limit.  Particularly, in the case of the $\pi^0$, which
has the smallest mass in the hadron spectrum, higher order corrections
to chiral anomaly is predicted to be small and can be calculated with
a sub-percent accuracy~\cite{Goity02,Ananth02,Ioffe07}.  As a result,
the precision measurement of $\pi^0\rightarrow \gamma\gamma$ decay
width is widely recognized as an important test of QCD.  On the other
hand, the system of $\pi^0$, $\eta$ and $\etap$ contains the
fundamental information on the effects of SU(3) and isospin symmetry
breaking by the light quark masses.  The two-photon decay widths of
$\eta$ and $\etap$ have a significant impact on the knowledge of the
quark-mass ratio $\displaystyle (m_d - m_u)/m_s$ \cite{Leutwyler96} and the
$\eta$-$\etap$ mixing angle.  Precision measurements of the two-photon
decay widths of these pseudoscalar mesons will provide a rich
experimental data set to understand QCD at low energies.

With the recent availabilities of high energy and high precision
intense photon tagging facilities~\cite{HallB-Tagger} together with
the novel developments in electromagnetic calorimetry it is feasible
to perform percent level differential cross section measurements of
light pseudoscalar mesons $\pi^0$, $\eta$ and $\etap$ on
nuclei~\cite{PrimEx-1}-\cite{JLAB12GeV}.  The two-photon decay widths
of these mesons can be extracted from these experiments, performed at
forward angles using the Primakoff method~\cite{Primakoff51}, which
assumes production of mesons in the Coulomb field of nuclei.  However,
the production process may also be manifested by exchange of vector
mesons having the same quantum numbers as the photon.  The angular
distributions of these two production mechanisms are realized
differently in the cross sections.  The Primakoff production is very
sharply peaked at forward angles (0.02 degree at $E_{\gamma} \sim $5
GeV) with practically vanishing strength at few degrees.  The strong
production is small under the Primakoff peak, but it begins to
dominate at few degrees.  The interference of these two coherent
amplitudes with its significant magnitude under the Primakoff peak,
makes the extraction of the decay widths quite difficult.  The full
theoretical description of this process, in addition to the above two
mechanisms, requires a correct treatment of the final state
interactions (FSI) of the produced mesons in nuclear matter, as well
as accounting for incoherent processes.  This is necessary to provide
a percent level extraction of the decay widths from the experimental
data set.  In this paper we present more general, and perhaps more
complete, theoretical descriptions of these photoproduction mechanisms
in nuclei.  We focus on the example of $\pi^0$ production keeping in
mind that the presented results can be extended to $\eta$ and $\etap$
production with appropriate replacements of the parameters and
cross sections.

\begin{figure}[!ht]
\begin{center}
 \includegraphics[scale=0.3]{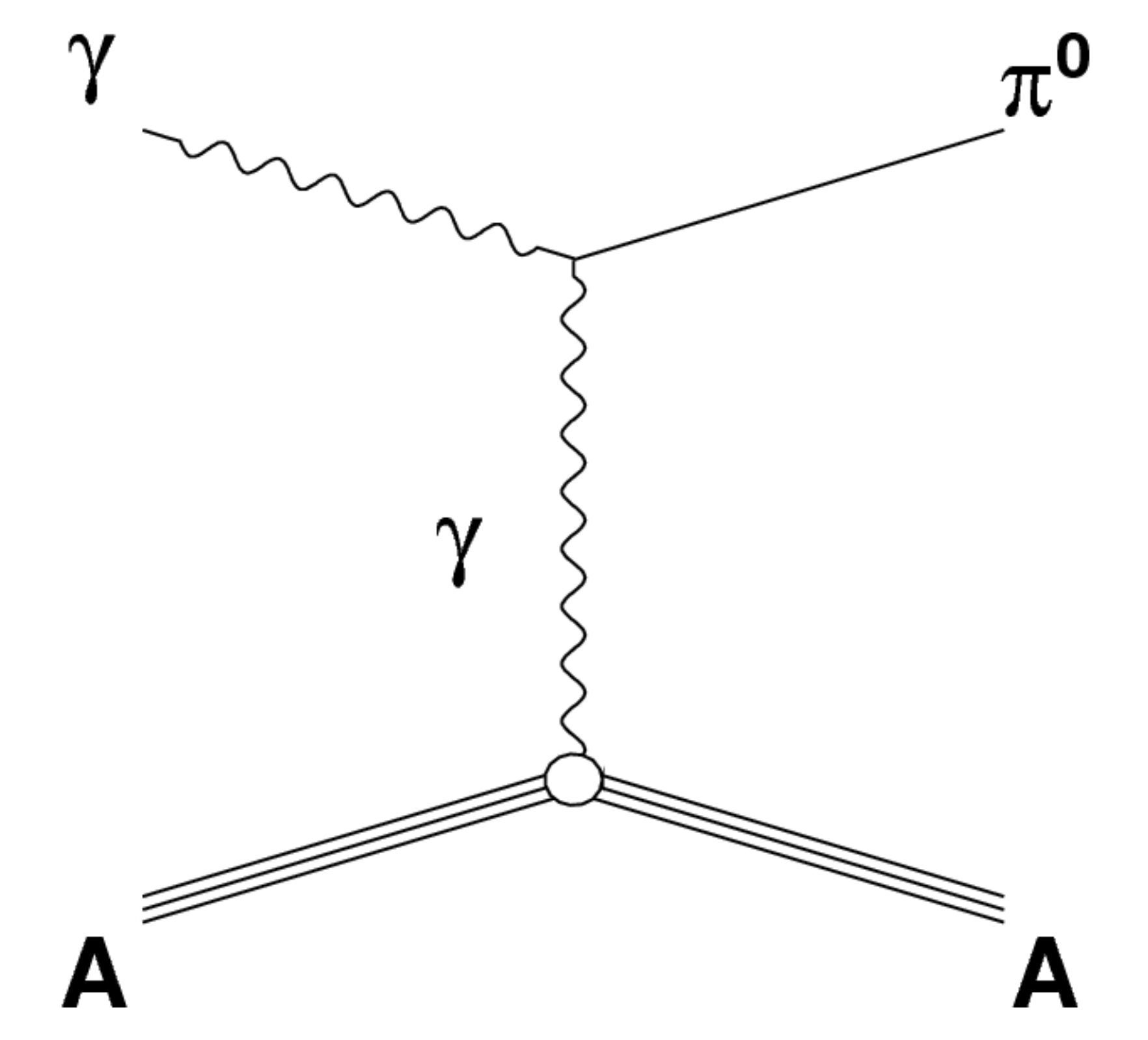}
 \vspace{-0.25cm}
 \caption{The pion photoproduction in the nuclear Coulomb field.}
 \label{fig1}
\end{center}
\end{figure}

In 1951 H. Primakoff~\cite{Primakoff51} first proposed to measure the
$\pi^0$ meson lifetime from their photoproduction in the Coulomb field
of a heavy nucleus (Fig.~\ref{fig1}).  With that we are considering
coherent photoproduction of pseudoscalar mesons on a nucleus at high
energies:
\ba
 \gamma + A \to Ps + A;~~~~ Ps=\pi^0,\eta,\eta^{\prime}
 \label{eq1}
\ea

\begin{figure}[!ht]
\begin{center}
\includegraphics[scale=0.3]{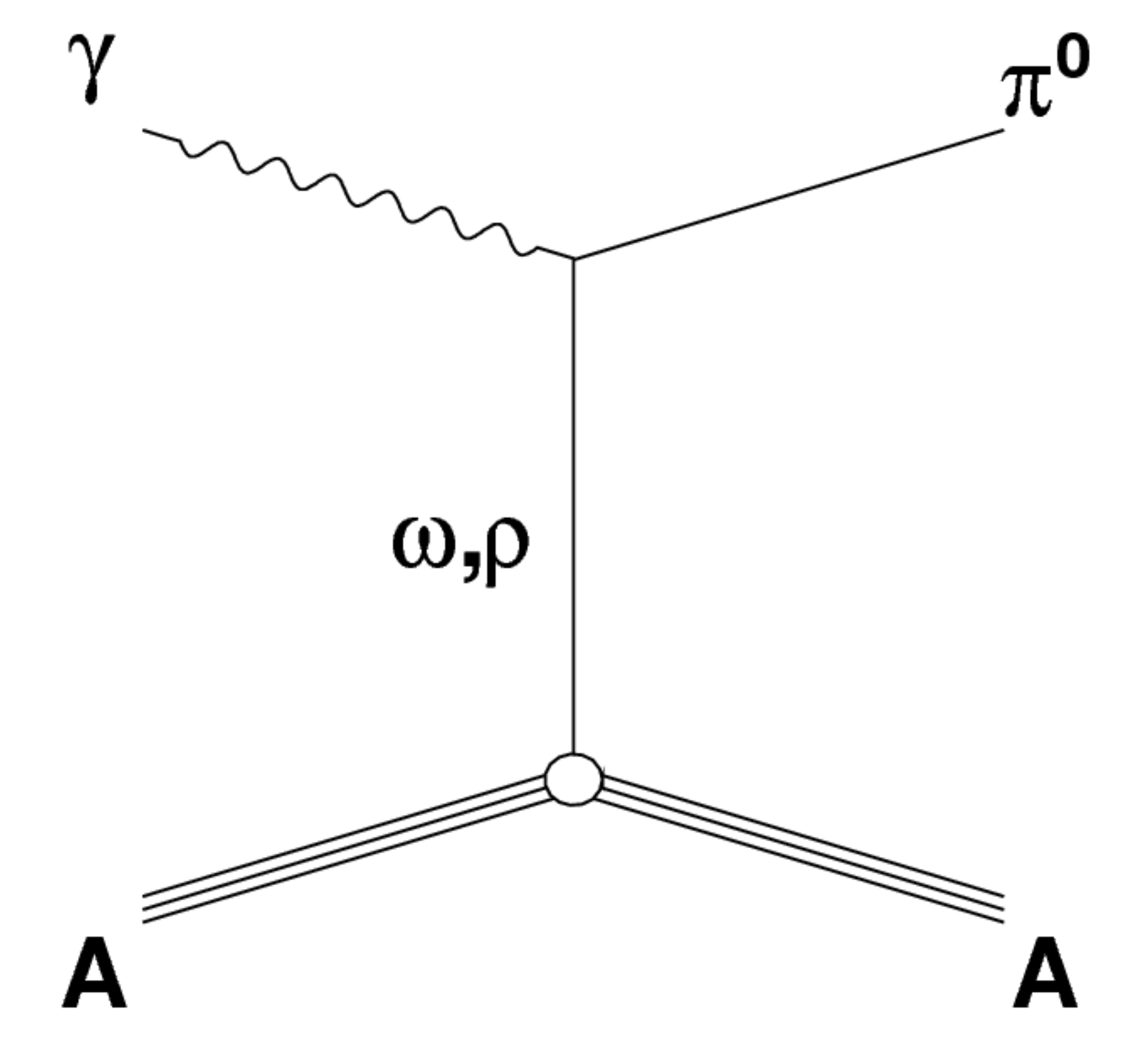}
 \vspace{-0.25cm}
 \caption{The pion photoproduction in the strong field of a nucleus.}
 \label{fig2}
\end{center}
\end{figure}

As mentioned above, the main challenge in the determination of pion
lifetime in this way is related to the presence of the strong
amplitude in the photoproduction process (Fig.~\ref{fig2}).  The full
amplitude of this coherent process can be described as a sum of the
Coulomb, $T_C$, and the strong, $T_S$, parts:
\ba
T=T_C+e^{i\varphi}T_S
 \label{eq2}
\ea
where $\varphi$ is the relative phase between the Coulomb and the
strong amplitudes.  Taking into the account the incoherent processes,
the differential cross section can be expressed as:
\ba
\frac{d\sigma}{d\Omega}=\frac{k^2}{\pi}\frac{d\sigma}{dt}
= ~\mid{T_C+e^{i\varphi}T_S}\mid^2
+ ~\frac{d\sigma_{inc}}{d\Omega}
\label{eq-2}
\ea
where $\displaystyle \frac{d\sigma_{inc}}{d\Omega}$ is the incoherent
cross section, {\em i.e.}, processes involving excitation or breakup
of the target nucleus.  Each of these amplitudes factorizes into a
photoproduction amplitude on a nucleon multiplied by a corresponding
form factor.  In addition, these form factors must be modified for the
final state interactions of the outgoing mesons with nucleons.
Interaction of incident high energy photons with nuclear matter gives
rise to a shadowing effect, which has to be considered as well.

\section{Photoproduction in the Coulomb field of a nucleus}

The effect of the pion Final State Interactions (FSI) in nuclei was
first discussed in detail by G. Morpurgo~\cite{Morpurgo64}, who
considered absorption of the produced pions in the strong nuclear field
using the Distorted Wave Approximation (DWA).  Calculations of
the electromagnetic and strong form factors in this work have been
done with the uniform nuclear density distribution $\rho(r)$.  Based
on these assumptions, part of the correction to the strong and
electromagnetic form factors, which takes into account pion absorption
in nuclei, is correctly obtained.  However, the effect of the pion
re-scattering to forward angles was not taken into account in this
work.  This effect is important for the precision extraction of the
decay widths since pions initially produced at modest angles can
re-scatter to small angles.  This effect was first considered by
G.F\"{a}ldt~\cite{Faldt72} in non-diffractive production processes on
nuclei in the framework of the Glauber theory of multiple
scattering~\cite{Glauber67}.
A general expression (Eq. 3.2 in~\cite{Faldt72}) for the electromagnetic
amplitude was obtained in this paper, but an analytically integrable
formula was derived for the case of equal absorption in a nucleus of the
incident and produced particles only (Eq. 3.4 in~\cite{Faldt72}),
which is not the case for the photoproduction processes.

\subsection{The electromagnetic form factor}
In a more general case, based on Glauber multiple scattering theory
and using the independent particle model for nucleons, the Coulomb
amplitude of the photoproduction of mesons on nuclei can be
expressed as:
\ba 
 T_C &= &Z\sqrt{8\alpha
 \Gamma}\left(\frac{\beta}{m_{\pi}}\right)^{3/2} \frac{k^2\sin{\theta}}
 {q^2+\Delta^2} F_{em}(q,\Delta)
 \label{eq4}
\ea 
where the electromagnetic form factor is given by~\cite{Gevorkyan-45}:
\ba
 F_{em}(q,\Delta)&=&\frac{q^2+\Delta^2}{q} \int J_1(q b)\frac{bd^2bdz}
 {(b^2+z^2)^{3/2}} e^{i\Delta z} \nonumber  \\ 
 &\times&
 \exp{\left(-\frac{\sigma'}{2}\int_{z}^{\infty}
 {\rho(b,z')dz'}\right)} \int_{0}^{\sqrt{b^2+z^2}}
 {x^2\rho(x)dx} 
 \label{eq5}
\ea
with the following notation: 
\ba 
 \sigma' &=&\sigma \left(1-i\frac{\Re f(0)}{\Im f(0)}\right)=\frac{4\pi}{ik}f(0)  \nonumber
\ea
Here $\sigma$ is the $\pi N$ total cross section, f(0) is the $\pi^0 N
\to \pi^0 N$ forward amplitude.  The invariant momentum transfer is
expressed through the two-dimensional transverse momentum $q=|\vec{q}|$
and longitudinal momentum $\Delta$:
$t=-q^2-\Delta^2=-4kp\sin^2\left(\frac{\theta}{2}\right)-\left(\frac{m_{\pi}^2}
{2k}\right)^2$  where $k=|\vec{k}|$  and $p=|\vec{p}|$ are the photon and pion momenta. 

In Eq.~\ref{eq5} $\rho(r)$ is the nuclear density, $J_1(x)$ is the
first order Bessel function~\cite{MathHBook64}.  The integration in
Eq.~\ref{eq5} goes over impact parameter $b$, which is the
two-dimensional vector in the plane perpendicular to the incident photon 
direction and the longitudinal coordinate $z$.

The Coulomb form factor, $F_{em}(q,\Delta)$, acquire an imaginary part
due to the longitudinal momentum transfer $\Delta$ and the presence of
the real part of the pion-nucleon elastic forward amplitude in the
absorption process.

The main properties of the Coulomb production are: it rises from zero
at zero degree angle; reaching its maximum at 
$\displaystyle q=\Delta=\frac{m_{\pi}^2}{2k}$ with the specific energy 
dependence at the peak of $\displaystyle \frac{d\sigma}{dt}\sim k^2$.  
These properties allow one to separate the Coulomb production
(Primakoff process) from the competing nuclear production in strong
field of nuclei, which peaks at relatively large production angles.

\begin{figure}[!ht]
\begin{center}
 \includegraphics[scale=0.7]{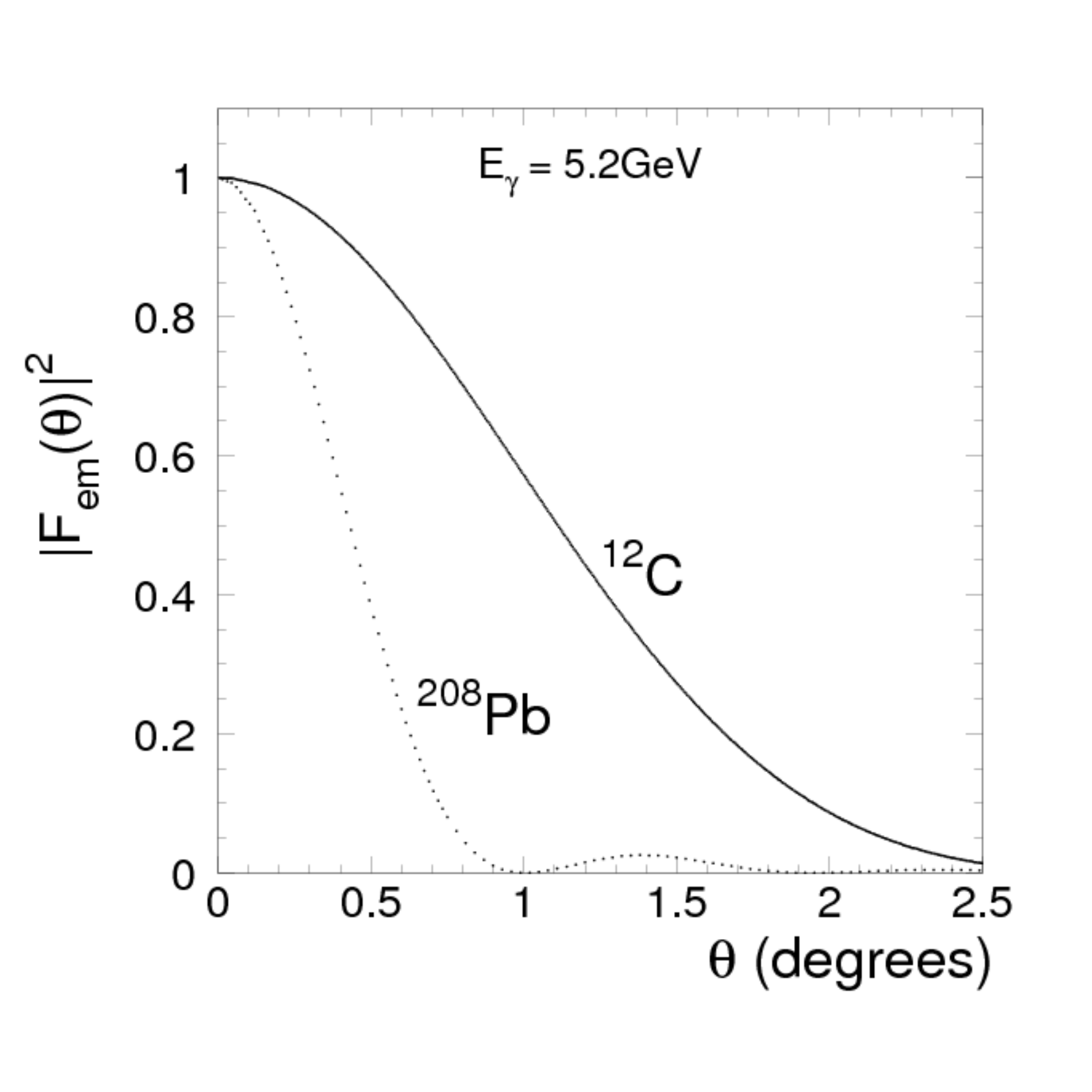}
 \vspace{-1.25cm}
 \caption{The square of the electromagnetic form factor for carbon (solid line)
  and lead (dotted line).}
 \label{fig3}
\end{center}
\end{figure}

For illustration, Fig.~\ref{fig3} shows the angular dependence of the
form factors calculated by Eq.~\ref{eq5} for carbon and lead nuclei at
5.2 GeV incident photon energy.  In this calculation the nuclear
density $\rho(r)$ was parametrized by Fourier-Bessel analysis from the
recent electron scattering data~\cite{DataTable74,Offerman91}.  As
evident, the final state interaction effects are more emphasized for
the heavier nucleus, though these effects are still very small in the
forward Primakoff region.

\subsection{Effect of light nuclei}

The photoproduction of mesons in the electromagnetic field of light
nuclei requires a special consideration.  Equation~\ref{eq5} was
obtained in the optical limit, which is valid for extended nuclear
matter (medium and heavy nuclei). Using multiple scattering
theory~\cite{Glauber67} we obtain the expression for the
electromagnetic form factor for light nuclei suitable for numerical
calculations~\cite{Gevorkyan-48}:
\vspace{-0.25cm}
\ba
 F_{em}(q)&=&\frac{q^2+\Delta^2}{q} \int J_1(q b)\frac{bd^2bdz}
  {(b^2+z^2)^{3/2}} e^{i\Delta z} [1-G(b,z)]^{A-1}\nn
 &\times& \int_{0}^{\sqrt{b^2+z^2}} {x^2\rho_{ch}(x)dx}\nn
 G(b,z)&=&\frac{f_s(0)}{ika_s}\int_{z}
 e^{-\frac{(\vec{b}-\vec{s}')^2}{2a_s}}\rho(s',z')d^2s'dz'\nn
 &=&
 \frac{\sigma'}{2a_s}\int_{z}
 e^{-\frac{ b^2+s'^2}{2a_s}}I_0\left(\frac{bs'}{a_s}\right)\rho(s',z')s'ds'dz'
\label{eq6}
\ea
Here $I_0(x)$ is the zero order Bessel function of imaginary
argument~\cite{MathHBook64}.  In deriving this expression, the
commonly used parametrization~\cite{Bauer78} for the elastic 
$\pi N\to \pi N$ amplitude $f_s(q)=f_s(0)exp(-a_s q^2/2)$ has been adopted.

For light nuclei, such as carbon, the nuclear density, $\rho(r)$,
corresponding to the harmonic oscillator potential well is widely
accepted in literature~\cite{Elton61}:
\ba
 \rho(r) =\frac{4}{\pi^{3/2}a_0^3}\left(1+\frac{A-4}{6}\frac{r^2}{a_0^2}\right)
 e^{-\left(\frac{r^2}{a_0^2}\right)} 
 \label{eq7}
\ea 
The nuclear charge distribution, $\rho_{ch}(r)$, is obtained by
convolution of the nuclear density (Eq.~\ref{eq7}) with the proton's
charge distribution.  For the latter we adopt the simple Gaussian
parametrization:
$\displaystyle \rho_p(r)=\frac{1}{\pi^{3/2}r_p^3}e^{-\left(\frac{r^2}{r_p^2}\right)}$.
With that, the nuclear charge density can be expressed by\footnote
{The nuclear density is normalized to the atomic number $\displaystyle \int\rho(r)d^3r=A$,
 whereas the charge density is normalized to the number of protons $Z$.}:
\ba
 \rho_{ch}(r)\hspace{-0.32cm}&=&\hspace{-0.27cm} \int d^3r'\rho(r')
 \rho_p(|\vec{r}-\vec{r}'|) \nn
 \hspace{-0.32cm}&=& \hspace{-0.27cm}
 \frac{2}{\pi^{3/2}(a_0^2+r_p^2)^{3/2}}
 \left[1+\frac{(Z-2)}{3}\left(\frac{3r_p^2}{2(a_0^2+r_p^2)}+\frac{a_0^2r^2}
 {(a_0^2+r_p^2)^2}\right)\right]e^{-\left(\frac{r^2}{r_p^2+a_0^2}\right)} \nn
 \label{eq8}
\ea

\begin{figure}[!ht]
\begin{center}
 \includegraphics[scale=0.7]{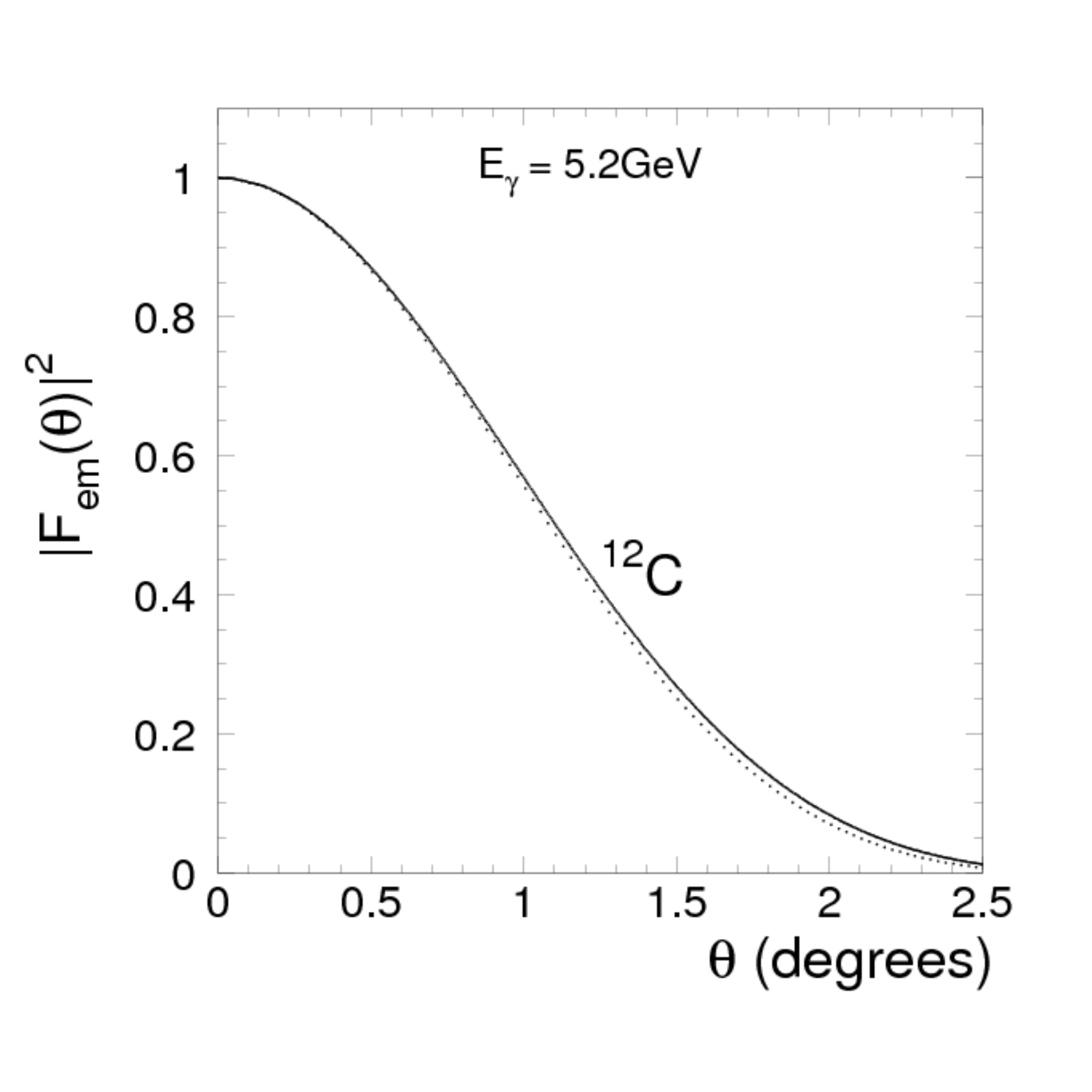}
 \vspace{-1.25cm}
 \caption{The square of electromagnetic form factor for carbon calculated by 
  expression (6)(dotted line) and (5) (solid  line).}
 \label{fig4}
\end{center}
\end{figure}

Figure~\ref{fig4} shows the difference between these two approaches on
the electromagnetic form factor calculations for carbon nucleus.  The
following parameters are used in these calculations: the proton radius
$r_p=0.8$ fm; oscillator parameter in Eq.~\ref{eq7}
$a_0=1.65$ fm.  It is seen that the form factor calculated by the
``light nuclei'' approach is falling faster at relatively large angles
than that based on the ``optical limit''.  Again, the difference
between these two methods in the Primakoff region is practically
negligible.

\subsection{Nuclear excitation by Coulomb exchange}

In addition to the coherent photoproduction in the Coulomb field,
production of pions with a nuclear collective excitation (for
instance, the giant dipole resonance) is also possible.  Such
processes were first discussed in~\cite{Glaser61} where for the
inelastic form factor the following approximation was obtained:
\ba
 |F_n(q)|^2\approx\frac{1.4N}{2m_pZAE_{av}}|t|
 \label{eq9}
\ea
where $m_p$ and $E_{av}$ are the proton mass and the average
excitation energy of the nucleus having $Z$ protons and $N$ neutrons.
The longitudinal momentum transfer in the case of $\pi^0$
photoproduction with nuclear collective excitation is much larger than
in the coherent photoproduction:
$\Delta_{in}=\Delta +E_{av}\gg \Delta$.
Using the fact that the ratio of the ``elastic'' to the ``inelastic''
cross sections of the $\pi^0$ photoproduction in the Coulomb field
can be estimated as:
\ba
 R=\frac{\frac{d\sigma_{in}}{d\Omega}}{\frac{d\sigma_{el}}{d\Omega}}
 \approx \frac{(q^2+\Delta^2)^2}{(q^2+\Delta_{in}^2)^2}\frac{|F_n(q)|^2}
 {|F(q)|^2}\approx\frac{1.4N}{2m_pZAE_{av}}\frac{(q^2+\Delta^2)^2}
 {(q^2+\Delta_{in}^2)}
 \label{eq10}
\ea

For the carbon nucleus the average collective excitation energy is on the level of 
$E_{av}\sim 20-25$ MeV which, using Eq.~\ref{eq10}, leads to $R\sim 10^{-7}$
in the Coulomb peak region ($\displaystyle q=\Delta=\frac{m_{\pi}^2}{2k}$).
Thus, the contribution from the nuclear collective excitations can be safely
neglected for the GeV and higher incident photon energies.

\section{Coherent photoproduction in strong field of nuclei}
\subsection{The strong amplitude $T_S$}

The main impact on the lifetime extraction from the measured
differential cross sections comes from our knowledge of the strong
amplitude $T_S$ in the coherent process of the reaction:
\ba
 \gamma + A \to \pi^0 + A
 \label{eq11}
\ea

In the Glauber theory of multiple scattering this coherent photoproduction amplitude 
is given by~\cite{Faldt72}:
\ba
 T_S(q,\Delta )&=& \frac{ik}{2\pi}\int e^{i\left(\vec{q} \cdot \vec{b}+\Delta z\right)}
 \Gamma_p(\vec{b}-\vec{s})\rho(\vec{s},z) \nn 
 &\times&
 \left[1-\int\Gamma_s(\vec{b}-\vec{s}')\rho(\vec{s}',z')d^2s' dz'\right]^{A-1}d^2bd^2sdz
 \label{eq12}
\ea
The two dimensional vectors $\vec{b}$ and $\vec{s}$ are the impact parameter
and the nucleon coordinate in the plane transverse to the incident photon
momentum; $z$ is the longitudinal coordinate of the nucleon in the nucleus.
The profile functions $\displaystyle \Gamma_{p,s}(\vec{b} - \vec{s})$ are
the two dimensional Fourier transforms of the non-spinflip elementary 
amplitudes for the pion photoproduction off the nucleon
$\displaystyle f_p=f(\gamma+N\to \pi^0+N)$ and elastic pion-nucleon scattering
$\displaystyle f_s=f(\pi+N\to\pi+N)$:
\ba 
 \Gamma_{p,s}(\vec b-\vec s)&=&\frac{1}{2\pi ik}
 \int e^{i\vec q \cdot (\vec b-\vec s)}f_{p,s}(q)d^2q
 \label{eq13}
\ea

Since the slope of elementary amplitude, $f_s(q)$, is typically much
less than the square of the nuclear radii for medium and heavy nuclei,
the nuclear density $\rho(r)$ is varying slower than the elastic
profile functions $\displaystyle \Gamma_s(\vec b-\vec s)$.  With this
approximation it is safe to take the $\rho(r)$ off from under the 
integral sign, which leads to:
\ba 
 \int\Gamma_s(\vec{b}-\vec{s})\rho(s,z)d^2sdz=
 \frac{\sigma'}{2}\int\rho(\vec{b},z)dz
 \label{eq14}
\ea

With special care the same procedure can also be applied for the production 
amplitudes~\cite{Faldt72}.

Isolating near zero angle part of the production amplitude 
$\displaystyle f_p(q) = (\vec{h} \cdot \vec{q}) \phi(q)$ 
($\displaystyle \vec{h}=\frac{[\vec{k}\times\vec{\epsilon}]}{k}$,
where $\epsilon$ is the photon polarization vector, and 
$\phi(0)\neq 0$) we get:
\ba
 \int\Gamma_p(\vec{b}-\vec{s})\rho(s,z)d^2sdz=
 \frac{2\pi}{k}\phi(0)\vec{h}\cdot\int\frac{\partial\rho(\vec{b},z)}
 {\partial\vec{b}}dz 
 \label{eq15}
\ea 
As a result, Eq.~\ref{eq12} can be written in the factorized form: 
\ba 
 T_S(q)&=&(\vec{h} \cdot \vec{q})\phi(0)F_{st}(q,\Delta) \nn 
 F_{st}(q,\Delta)&=&-\frac{2\pi}{q}\int J_1(qb)\frac{\partial\rho(b,z)}
 {\partial b} b db dz e^{i\Delta z}\exp\left(-\frac{\sigma'}{2}
 \int_{z}^{\infty}\rho(b,z')dz'\right) \nn
 \label{eq16}
\ea

The strong form factor, $F_{st}(q,\Delta)$, can be expressed as a sum of
two terms, as done in~\cite{Faldt72}:
\ba 
 &F_{st}(q,\Delta)&=\int e^{i\vec{q} \cdot \vec{b}}\rho(b,z)d^2bdz
 e^{i\Delta z}\exp\left(-\frac{\sigma' }{2}
 \int_{z}^{\infty}\rho(b,z')dz'\right) \nn
 &-& 
 \hspace{-0.75cm} \frac{\pi\sigma'}{q} \hspace{-0.10cm}\int J_1(qb)\rho(b,z_1)
 \frac{\partial\rho(b,z_2)}{\partial b} b db dz_1dz_2  e^{i\Delta z_1}
 \exp\left(-\frac{\sigma'}{2}\hspace{-0.10cm}\int_{z_1}^{\infty}
 \hspace{-0.20cm}\rho(b,z')dz'\right) \nn
 \label{eq17}
\ea
The first term is the usual nuclear form factor and the second one is
a correction first introduced by F\"{a}ldt~\cite{Faldt72}.  The
contribution from the second term is positive (since $\displaystyle
\frac{\partial\rho(b,z)}{\partial b}<0$) and can be interpreted as a
result of the final state interaction of the pions produced in the
coherent process at nonzero angles which after multiple scattering
come to forward direction.  Note that this correction was
obtained~\cite{Faldt72} as in the case of electromagnetic part under
the assumption that the cross sections for the incident and produced
particles are equal.  We have considered a more general case for the
photoproduction reactions and the results are discussed in the next
sections.

\subsection{Photon shadowing in nuclei}

Real photons at high energies are shadowed in nuclei~\cite{Bauer78}.
Photon shadowing in pion photoproduction is a result of the two-step
process~\cite{Gottfried69}: the initial photon produces a vector
meson in the nucleus, which consequently produces the pseudoscalar
meson on another nucleon of the same nucleus.  The main contribution
to this process comes from the $\rho$ mesons, as the cross section for
$\rho$ photoproduction from a nucleon is almost one order of magnitude
greater than that for $\omega$ production.  In addition, the reaction
$\omega+N\to \pi^0+N$ is mainly caused by isospin one exchange
($\rho$-exchange) and the production amplitudes on protons and
neutrons have different signs.  Since the nuclear coherent amplitude
is a sum over the elementary photoproduction amplitudes the omega
contribution gets an additional suppression.

Using multiple scattering techniques, one can obtain the contribution
from intermediate $\rho$ channel to the strong form factor: 
\ba
 F_I(q)=-\frac{\pi\sigma'}{q} \int J_1(q b)\rho(b,z_1)\frac{\partial
 \rho(b,z_2)}{\partial b} \theta(z_2-z_1)bdbdz_1dz_2 \nn
 \times ~e^{i\Delta_{\rho}(z_1-z_2)+i\Delta z_2} \exp\left(-\frac{\sigma'}{2}
 \int_{z_1}^{\infty}\rho(b,z')dz\right)
 \label{eq18}
\ea
where $\displaystyle \Delta_{\rho}=\frac{m_{\rho}^2}{2E}$ is the
longitudinal momentum transfer in the $\rho$ meson photoproduction off
the nucleon.  The strong amplitude accounting for the photon shadowing
reads:
\ba
 T_S(q)&=&(\vec h \cdot \vec q )\phi(0)(F_{st}-wF_I) \nn
     w &=&\frac{f(\gamma N\to \rho N) f(\rho N\to \pi N)}{f(\rho N\to \rho N)
       f(\gamma N\to \pi N)}
 \label{eq19}
\ea
where the range of the shadowing parameter $w$ is between zero (no shadowing) 
and one (Vector Dominance Model).

Equation~\ref{eq18} and F\"{a}ldt's correction (second term in
Eq.~\ref{eq17}) is very similar and only difference is in the energy
dependence through the longitudinal momentum transfer.  It is
interesting to mention here that at high energies where
$\Delta=\Delta_{\rho}=0$, assuming a validity of the naive Vector
Dominance Model ($w=1$), the above mentioned two terms cancel each
other.  Thus, the strong form factor is determined only by the first
term in Eq.~\ref{eq17}.  In this limit the integration over $z$ in
Eq.~\ref{eq17} can be done analytically:
\ba
 F_{st}(q)=\frac{2}{\sigma'}\int d^2be^{i\vec{q} \cdot \vec{b}}
 \left[1-\exp\left(-\frac{\sigma'}{2}\int\rho(b,z')dz\right)\right]
 \label{eq20}
\ea
This expression looks very similar to the one for coherent production
of pions by $\rho$ mesons, a fact well known for the diffractive
processes~\cite{Bauer78}.

\begin{figure}[!ht]
\begin{center}
 \includegraphics[scale=0.7]{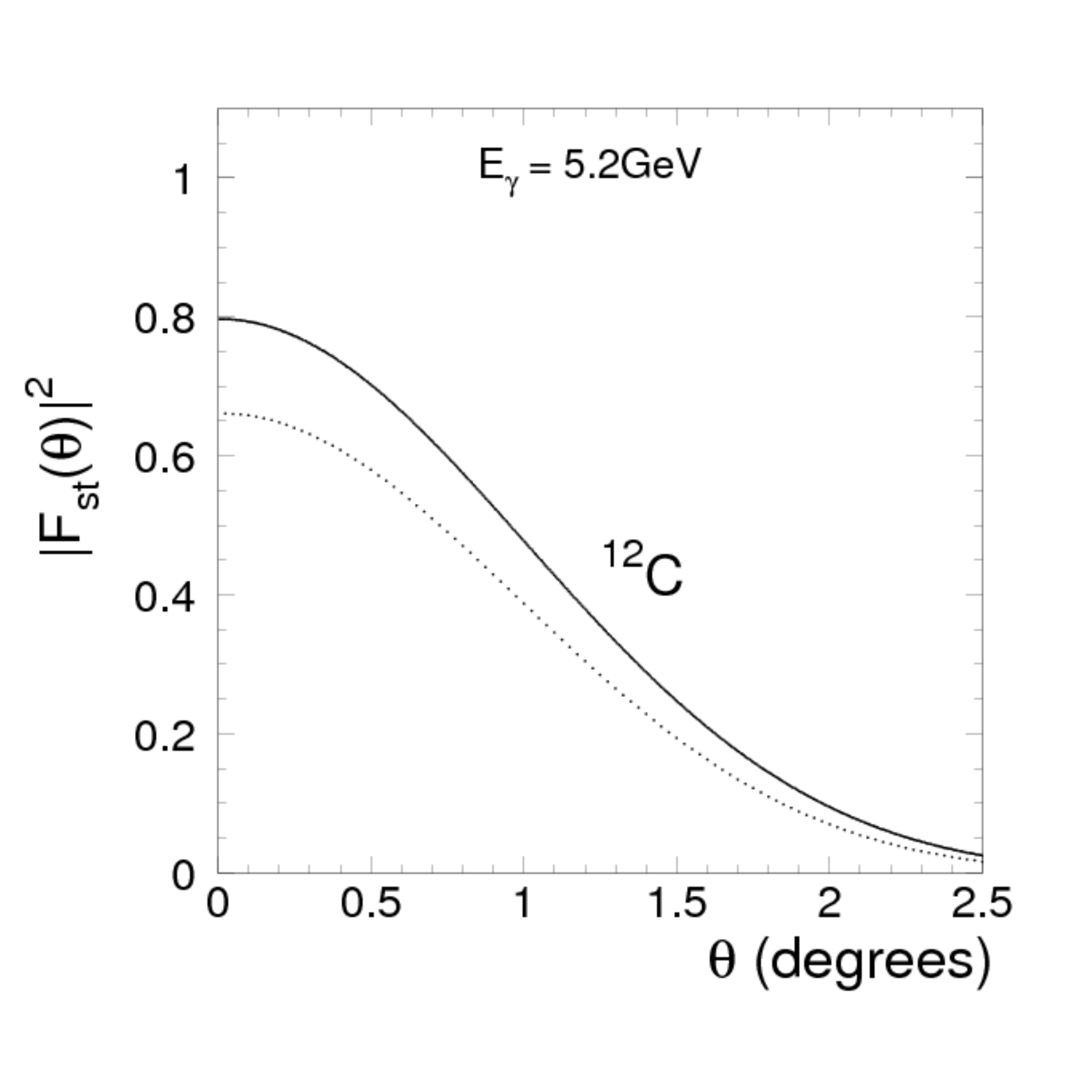}
 \vspace{-1.25cm}
 \caption{The square of the strong form factor for carbon without shadowing
  ($w=0$, solid line) and with maximal photon shadowing ($w=1$, dotted line).}
 \label{fig5}
 \end{center}
\end{figure}

\begin{figure}[!ht]
\begin{center}
 \includegraphics[scale=0.7]{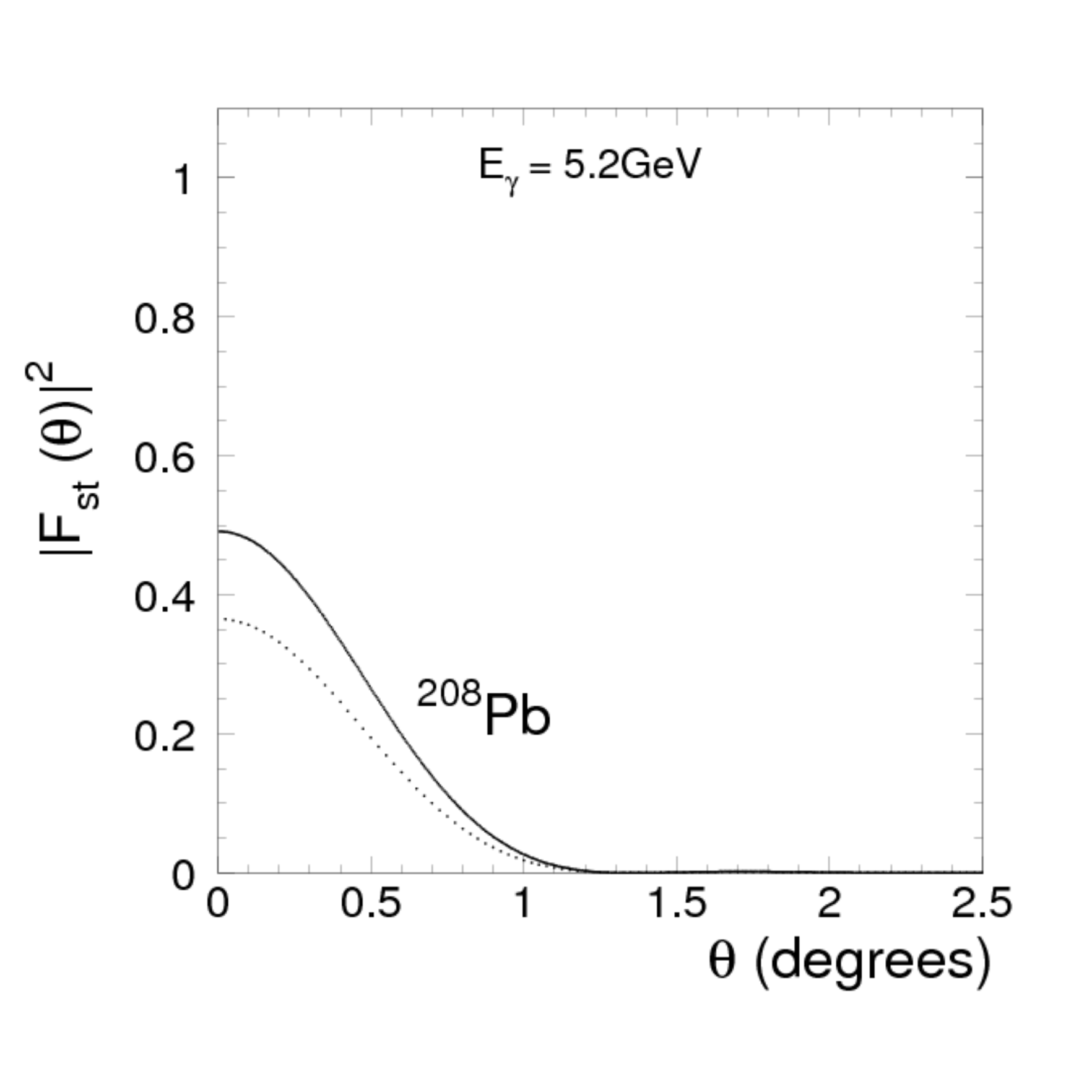}
 \vspace{-1.25cm}
 \caption{The square of the strong form factor for lead  without shadowing
  ($w=0$, solid line) and with maximal photon shadowing ($w=1$, dotted line)}
 \label{fig6}
 \end{center}
\end{figure}

The effect of photon shadowing on the strong form factor in
Eq.~\ref{eq19} is demonstrated in Figs.~\ref{fig5} and \ref{fig6} for
carbon and lead nuclei, calculated from Eqs.~\ref{eq17}-\ref{eq19} for
two extreme values of the shadowing parameter: $w=0$ (no shadowing)
and $w=1$ (naive VDM).  As it is seen, the results for both nuclei are
strongly dependent on the value of $w$.

\subsection{The two-step contribution in light nuclei}

The optical limit approximations, usually used in the literature for
these calculations, can potentially lead to a correction for light
nuclei, which can be critical for the precision extraction of decay
width from the experimental cross sections.

For light nuclei we parametrize the elementary production amplitudes
in the following way:
\ba 
 f_p&=&\phi(0)(\vec{h} \cdot \vec{q})
 e^{-\frac{a_p q^2}{2}} \nn
 f_s &=&f_s(0) e^{-\frac{a_s q^2}{2}}
 \label{eq21}
\ea
Here $\phi(0), f_s(0)$ and $a_p, a_s$ are the forward elementary
amplitudes and their relevant slopes. With these parametrizations  
one gets:
\ba
 \Gamma_p(\vec{b}-\vec{s})&=&\frac{\vec{h} \cdot (\vec{b}-\vec{s})}{ka_p^2}\phi(0)
 e^{-\frac{(\vec{b}-\vec{s})^2}{2a_p}} \nn
 \Gamma_s(\vec{b}-\vec{s})&=&\frac{f_s(0)}{ika_s}
 e^{-\frac{(\vec{b}-\vec{s})^2}{2a_s}}
 \label{eq22}
\ea
Substituting Eq.~\ref{eq22} into Eq.~\ref{eq12} we obtain:
\ba
 T_S(q,\Delta)&=& (\vec{h} \cdot \vec{q}) \phi(0)F_{st}(q,\Delta) \nn
 F_{st}(q,\Delta)&=&
 \frac{2\pi}{qa_p^2}\int J_1(qb) 
 \left[bI_0\left(\frac{bs}{a_p}\right) - sI_1\left(\frac{bs}{a_p}\right)\right] \nn
 &\times&e^{i\Delta z}e^{-\frac{b^2+s^2}{2a_p}}\rho(s,z)
 \left[1-G(b,z)\right]^{A-1} bdbsdsdz \nn
 G(b,z)&=&\frac{f_s(0)}{ika_s}\int_{z}
 e^{-\frac{(\vec{b} -\vec{s}')^2}{2a_s}}\rho(s',z')d^2s'dz' \nn 
    &=& \frac{\sigma'}{2a_s}\int_{z}
 e^{-\frac{ b^2+s'^2}{2a_s}}I_0\left(\frac{bs'}{a_s}\right)\rho(s',z')s'ds'dz'
 \label{eq23}
\ea
The amplitude relevant to the two-step process $\gamma\to\rho\to\pi^0$
in multiple scattering theory is given by:
 \ba
 T_I(q)&=&\frac{ik}{2\pi}\frac{A-1}{A} \int e^{i\vec{q} \cdot \vec{b}}d^2b
 \Gamma_{\gamma \rho}(\vec{b}-\vec{s}_1)\Gamma_{\rho \pi}(\vec{b}-\vec{s}_2)
 \rho(s_1,z_1)\rho(s_2,z_2) \nn
 &\times& \theta(z_2-z_1)e^{i\Delta_{\rho}(z_1-z_2)+i\Delta z_2}
 \left[1-G(b,z_1)\right]^{A-2} d^2s_1d^2s_2dz_1dz_2 
 \label{eq24}
\ea
where $\displaystyle \Delta_{\rho}=\frac{m_{\rho}^2}{2E}$ is the longitudinal momentum
transfer in the elementary reaction $\displaystyle \rho +N\to \pi^0+N$.
Using Eq.~\ref{eq22} the relevant form factor can be expressed in a 
form convenient for numerical integration:
\ba
 F_{I}(q,\Delta_{\rho},\Delta)&=&\frac{A-1}{A}\frac{\pi\sigma'}{qa_sa_p^2}
 \int J_1(q b)I_0\left(\frac{bs_2}{a_s}\right)
 \left[bI_0\left(\frac{bs_1}{a_p}\right) 
 -s_1I_1\left(\frac{bs_1}{a_p}\right)\right] \nn
 &\times& \hspace{-0.25cm} \theta(z_2-z_1)
 \rho(s_1,z_1) \rho(s_2,z_2)
 e^{-\frac{(a_p+a_s)b^2}{2a_pa_s}}
 e^{-\frac{s_1^2}{2a_p}-\frac{s_2^2}{2a_s}}
 e^{i\Delta_{\rho}(z_1-z_2)+i\Delta z_2} \nn
 &\times& \hspace{-0.25cm} \left[1-G(b,z_1)\right]^{A-2}
 bdbs_1ds_1s_2ds_2dz_1dz_2 
 \label{eq25}
\ea

\begin{figure}[!ht]
\begin{center}
 \includegraphics[scale=0.7]{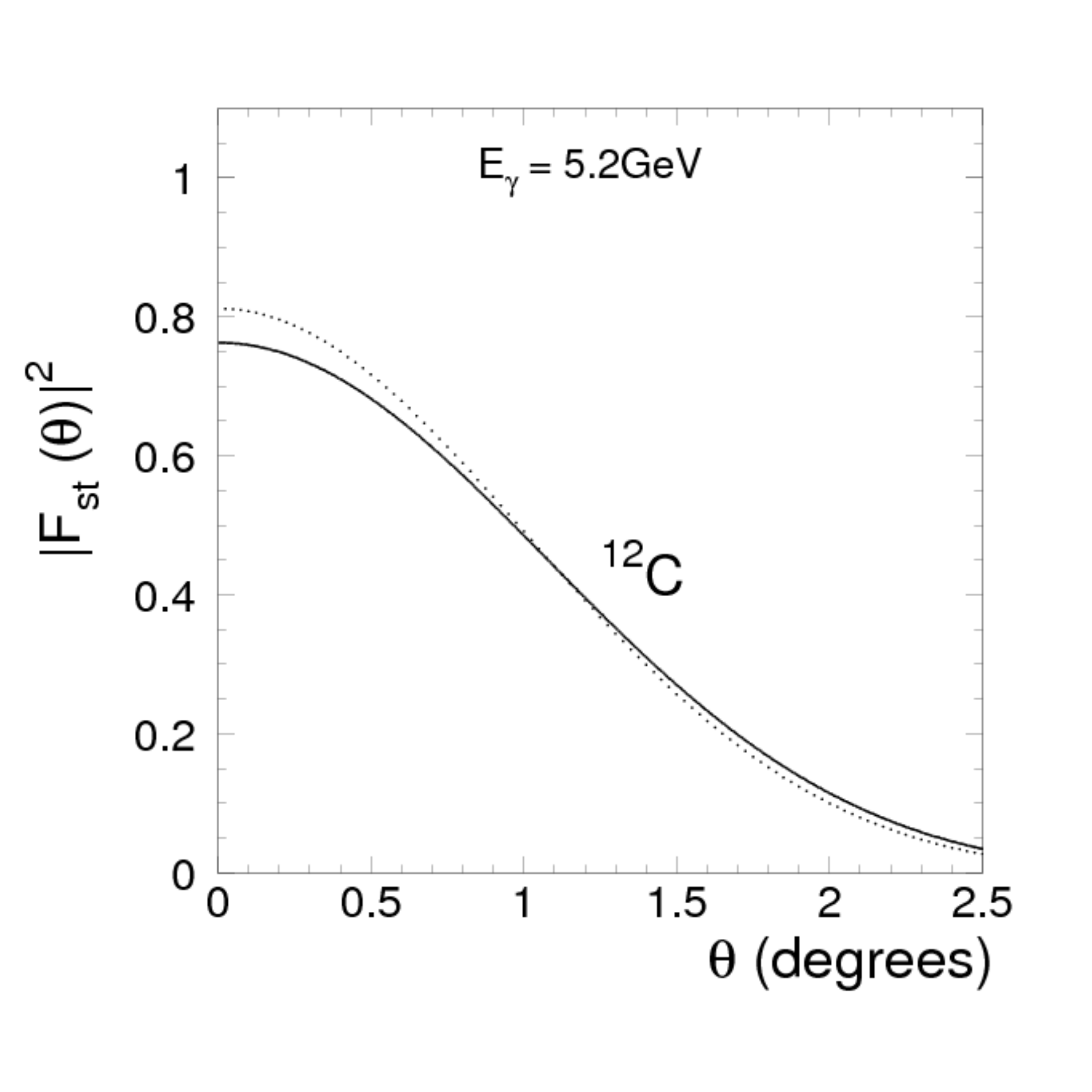}
 \vspace{-1.25cm}
 \caption{The square of the strong form factor for carbon nucleus calculated
  with (solid line) and without (dotted line) optical model approximations.}
 \label{fig7}
 \end{center}
\end{figure}

In Fig.~\ref{fig7}, the square of the strong form factor for carbon
nucleus including the intermediate channel corrections are plotted for
two different cases:
(a) the solid line is calculated by Eqs.~\ref{eq17}-\ref{eq19} with
optical model approximations;
(b) the dotted line is calculated by multiple scattering theory for
light nuclei without optical model approximations (using
Eqs.~\ref{eq23}-\ref{eq25}).  The oscillator type of nuclear density
(Eq.~\ref{eq7}) and the shadowing parameter $w=0.25$
(\cite{Meyer72,Boyarski69}) are used in both calculations.  As it is
seen, the fall of the form factor calculated without the optical model
approximations (dotted line) is sizeably faster, and therefore
important for the precision extraction of the decay widths.  A similar
behavior was observed for the electromagnetic form factor shown in
Fig.~\ref{fig4}.

\section{Incoherent photoproduction}

\begin{figure}[!ht]
\begin{center}
\includegraphics[scale=0.65]{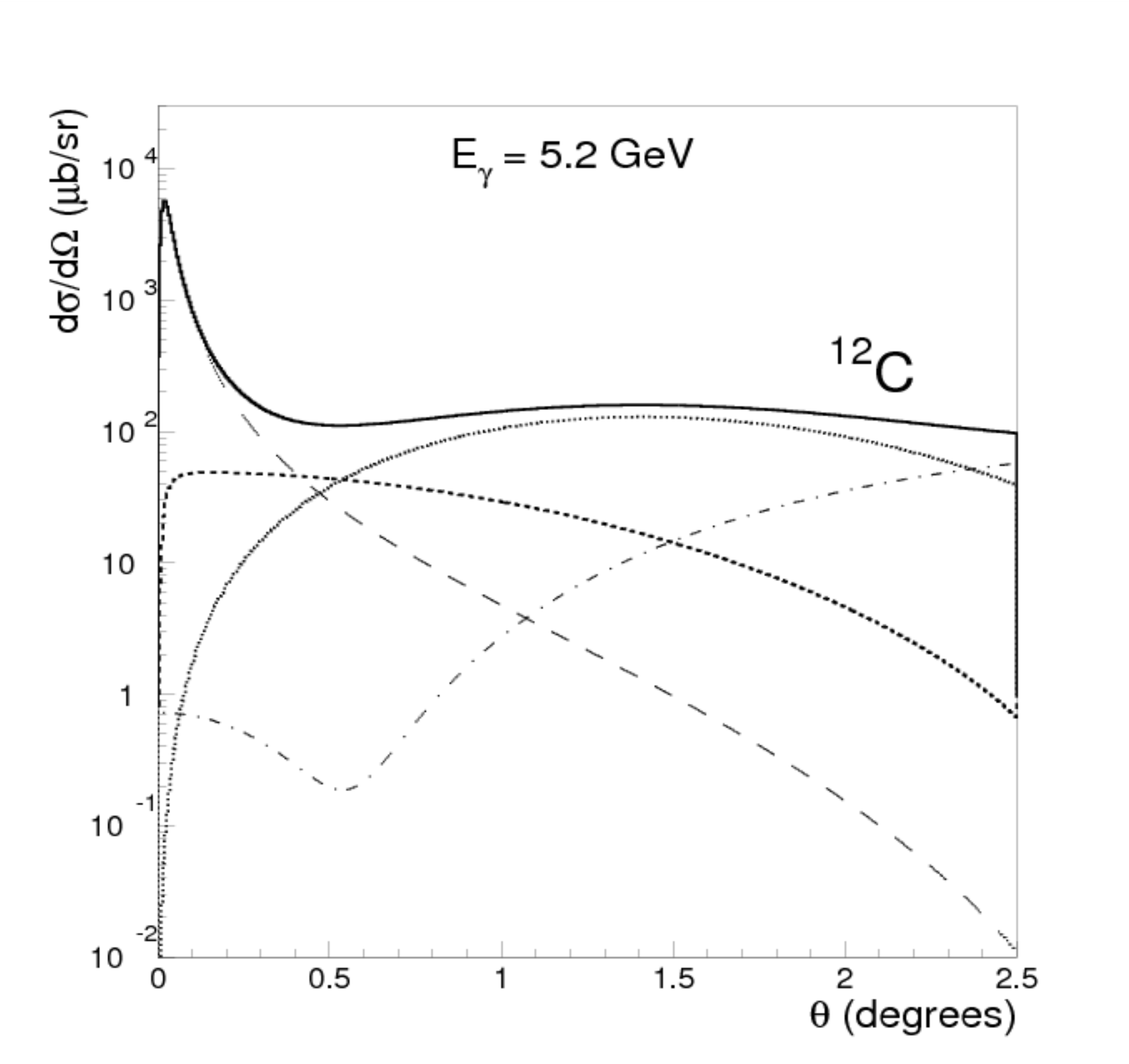}
 \vspace{-0.50cm}
 \caption{Differential  cross section for $ \pi^0 $ photoproduction off 
  carbon nuclei as a function  of the production angle in the lab system. 
  The long-dashed line shows the electromagnetic (Primakoff) contribution;
  the dotted line is the strong part; 
  the dash-dotted line is the incoherent cross section;
  the long-dashed line is the interference between Primakoff and strong amplitudes;
  the solid line is the full cross section.}
 \label{fig8}
\end{center}
\end{figure}

Incoherent meson photoproduction is a production with the excitation
or a breakup of the target nucleus:
\ba 
 \gamma+A\to \pi^0+A'
 \label{eq26}
\ea
%
The general expression for the incoherent cross section established in
the literature~\cite{Engelbrecht64,Bellettini70} is given by:
\ba
 \frac{d\sigma_{inc}}{d\Omega}=
 \frac{d\sigma_0}{d\Omega}(q)N(0,\sigma)\left[1-G(t)\right]
 \label{eq27}
\ea
where $\displaystyle \frac{d\sigma_0}{d\Omega}$ is the elementary
cross section on the nucleon $\gamma+N\to \pi^0+N$, and
\ba
 N(0,\sigma)=\int \frac{1-e^{-\sigma  T(b)}}{\sigma}d^2b
 \label{eq28}
\ea
Here $\displaystyle T(b)=\int \rho(b,z)dz$.  The factor
$\displaystyle \left[1-G(q)\right]$ takes into account the suppression
of pions produced at small angles due to Pauli exclusion
principle~\cite{Engelbrecht64} and goes to zero at small angles.  But
as was shown for the case of the proton scattering on
nuclei~\cite{Glauber70}, the multiple scattering leads to the
incoherent cross section, which is different from zero at zero angles.
The same effects should also be in the pseudoscalar mesons
photoproduction.  Assuming that the meson photoproduction cross
section on the nucleon is completely determined by the spin-nonflip
amplitude, one can parametrize the differential cross section on the
nucleon as\footnote{Numerically this parametrization at small 
momentum transfer coincides with the predictions for $\pi^0$ photoproduction 
cross section on proton obtained~\cite{Sibirtsev09} 
in the framework of improved Regge theory}
\ba
 \frac{d\sigma_0}{d\Omega}=c_p q^2e^{-a_pq^2}
 \label{eq29}
\ea
Using this parametrization it can be shown\footnote{The derivation
and detailed discussion of the incoherent production at small angles
will be given elsewhere.} that the incoherent cross section of the
process (Eq.~\ref{eq26}) can be represented as
\ba
 \frac{d\sigma_{inc}}{d\Omega}&=&
 \frac{d\sigma_0}{d\Omega}(q)\left[N(0,\sigma)-\frac{|F_{st}(q)|^2}{A}\right]
 +c_p\xi^2e^{-a_pq^2}\nn
 \xi^2&=&|\frac{\sigma'}{2}|^2\int\rho(b,z) 
 |\frac{\partial T(b,z)}{\partial b}|^2\exp \left(-\sigma T(b,z)\right)d^2bdz \nn
 \label{eq30}
\ea
Here $F_{st}(q)$ is the strong form factor of the nucleus
(Eq.~\ref{eq17}). One can easily see that only in the case when
absorption is absent ($\sigma=0$) the factorization similar to
Eq.~\ref{eq27} takes place.\\ 
In Fig.~\ref{fig8} we plot the
differential cross section for the $\pi^0$ photoproduction off carbon
nucleus $vs$ laboratory angle. The different curves gives the
contribution of relevant mechanism discussed above. All calculations
have been done using the expressions obtained in the present work.
The spin-nonflip $\pi^0$ photoproduction amplitude on the nucleon
was parametrized as ~\cite{Cornell74,SLAC71} $\displaystyle
f(q)=10\left(\frac{k}{k_0}\right)^{1.2} \sin{\theta}e^{i\varphi}$
where $\varphi\sim$ 1 radian and $k_0=1$ GeV. The differential cross
section of the $\pi^0$ photoproduction on the proton was taken in
accordance with existing experimental data~\cite{Braunschweig70} at
forward angles.
 
\section{Summary}

We have extended the theoretical treatment of forward production of
pseudoscalar mesons off nuclei described in the literature in
connection with the extraction of their radiative decay widths.  Based
on the Glauber theory of multiple scattering we have derived complete
analytical formulas for the electromagnetic and strong form factors
specific for the photoproduction of pseudoscalar mesons off both light
and heavy nuclei, valid for realistic charge and matter distributions.
The electromagnetic form factor for the non-diffractive
photoproduction processes is derived for the first time.  Special
attention is paid to light nuclei, since with the increasing photon
beam energies the precision decay width extractions is becoming more
feasible from the light targets.  We have included in our results the
difference between the charge and matter distributions for the light
nuclei.  The photon shadowing effect in the reactions under
consideration is correctly treated for the first time.  The impact of
this effect on the $\pi^0$ meson production in the strong field of
both light and heavy nuclei are calculated.  An expression for the
incoherent photoproduction cross sections of the pseudoscalar mesons
at forward angles is derived, which correctly takes into account the
mesons' final state interactions and the exclusion principle.  Using the
obtained expressions we calculate the differential cross section of
the $\pi^0$ meson photoproduction off carbon nucleus and the
contributions of different mechanisms considered above.  All these give
a good challenge to extract the lifetime of pseudoscalar mesons from
forthcoming experimental data with high precision.

\vspace{0.75cm}
\noindent
{\Large \bf Acknowledgements}

\vspace{0.25cm}
 This work was inspired by the PrimEx experiment at Jefferson Lab to measure 
 the $\pi^0$ lifetime via the Primakoff effect with high precision.  It was 
 partially supported by the National Science Foundation grants PHY-0245407
 and PHY-0555524.

\end{document}